\documentclass[aps,prl,twocolumn,groupedaddress]{revtex4}

\usepackage{graphicx}
\usepackage{bm}

\begin{document}

\title{Planar metamaterial with transmission and reflection that depend on the direction of incidence}



\author{E. Plum} \email[Email: ]{erp@orc.soton.ac.uk} \affiliation{Optoelectronics Research Centre, University of Southampton, SO17 1BJ, UK}

\author{V. A. Fedotov}
\affiliation{Optoelectronics Research Centre, University of
Southampton, SO17 1BJ, UK}

\author{N. I. Zheludev}
\homepage[Homepage:]{www.nanophotonics.org.uk/niz/}
\affiliation{Optoelectronics Research Centre, University of
Southampton, SO17 1BJ, UK}

\date{\today}

\begin{abstract}
We report that normal incidence reflection and transmission of
circularly polarized electromagnetic waves from and through planar
split-ring metamaterials with chiral symmetry breaking depends on
the incidence direction and handedness of circular polarization. The
effect has a resonant nature and is linked to the lack of
mirror-symmetry in the metamaterial pattern leading to a
polarization-sensitive excitation of electric and magnetic dipolar
responses in the meta-molecules.
\end{abstract}




\maketitle

Directional asymmetry of transmission and reflection is normally
associated with the presence of a static magnetization of the medium
which breaks the reciprocity of the light-matter interaction. Here
the optical Faraday effect is the most important example. However,
recently it was understood that asymmetric transmission is possible
without breaking reciprocity in a lossy medium if polarization
conversion is involved. This has been seen for electromagnetic waves
propagating through planar chiral metamaterial structures
\cite{PRL_Fedotov_2006_AsymmetricTransmissionMW,
NanoLett_Fedotov_2007_AsymmetricTransmission,
NanoLett_Schwanecke_2008_AsymTrans}. Several other ideas using
chirality in asymmetric and polarization sensitive devices have
recently been suggested \cite{OptExp_Ebbesen_2008_AsymTrans,
MetaCongress_Zhukovsky_2008_EllDichroism,
Science_Genack_ChiralFiberGratings,
APL_Shvets_2006_ChiralWaveguide,hwang-2005-eta}. In this Letter we
demonstrate a type of metamaterial that shows strong \emph{resonant}
asymmetric transmission at normal incidence and we report that this
is accompanied by \emph{asymmetric reflection}: transmission and
reflection of circularly polarized light depend on the direction of
incidence and on the handedness of the incident circular
polarization state.

\begin{figure}[t!]
\includegraphics[width=85mm]{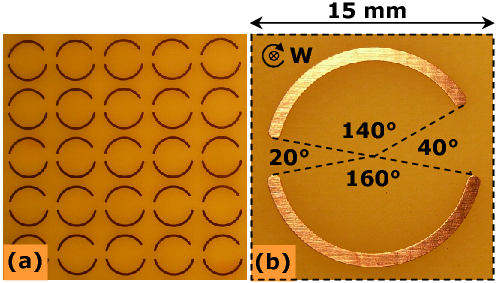}
\caption{\label{fig-structure}(Color online) (a) Front side of a
section of planar metamaterial formed by a square array of 2D-chiral
asymmetrically split rings. (b) Unit cell of the metamaterial. The
twist vector \textbf{W}, associated with the chirality of the unit
cell, points away from the reader indicating an overall clockwise
twist.}
\end{figure}

The effect has been observed in a novel type of planar metamaterials
based on asymmetrically split rings (ASR) supporting high-Q
trapped-mode resonances of collective nature
\cite{PRL_Fedotov_2007_TrappedModes,
arXiv_Papasimakis_2008_Coherent}. To see the asymmetric effects we
modified a previously used design by introducing an asymmetry in
both arcs and gaps so that the resulting metamaterial has no line of
mirror symmetry and thus is 2D-chiral: a ``twist vector" \textbf{W}
governed by the corkscrew law (rotation from small gap towards large
gap along the short arc) may be associated with the handedness
(twist) of the pattern (see Fig.~\ref{fig-structure}b). In our
structures the metal rings had a radius of 6~mm, a width of 0.8~mm
and were split to create two arcs of $160^\circ$ and $140^\circ$
with gaps corresponding to $30^\circ$ and $10^\circ$. They were
etched from $35~\mu\text{m}$ copper cladding covering 1.6~mm thick
FR4 PCB substrate ($\varepsilon \simeq 4.5$). The metamaterial was
formed by a regular array of these rings with an overall size of
approximately $220~\times~220~\text{mm}^2$ and a square unit cell of
$15~\times~15~\text{mm}^2$ (see Fig.~\ref{fig-structure}a).

Transmission and reflection properties of the metamaterial were
studied in an anechoic chamber in the 3.0 - 9.0~GHz spectral range
for waves normally incident on both ``front" and ``back" of the
structure using two linearly polarized broadband horn antennas
(Schwarzbeck BBHA 9120D) and a vector network analyzer (Agilent
E8364B). Note that waves incident from the material's front and back
propagate parallel and antiparallel to \textbf{W} respectively. The
complex circular transmission and reflection matrices, defined as
$E_i=t_{ij}E^{0}_j$ and $E_i=r_{ij}E^{0}_j$, where the indices $i,
j$ denote either right- or left-handed circularly polarized
components (RCP, + or LCP, -), were calculated directly from the
measured transmission and reflection matrices for orthogonal linear
polarizations. Intensities of the corresponding transmitted,
reflected and converted components for incident RCP and LCP waves
were calculated as $T_{ij}=|t_{ij}|^2$ and $R_{ij}=|r_{ij}|^2$.

\begin{figure}[t!]
\includegraphics[width=85mm]{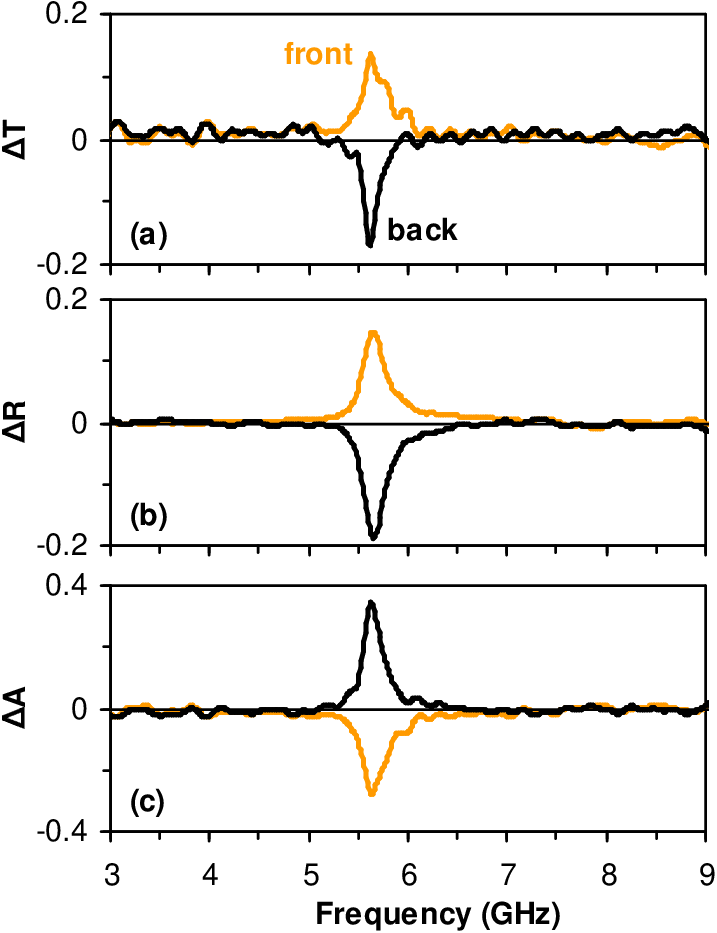}
\caption{\label{fig-TRA} (Color online) Difference in total
transmission (a), reflection (b) and absorbtion (c) of RCP- and
LCP-waves normally incident on the front (orange) and back (black)
of the chiral ASR-metamaterial.}
\end{figure}

We found that total transmission through the metamaterial (as it
would be measured with a polarization insensitive detector), defined
as $T_{j}=T_{jj} + T_{ij}$, was different for circularly polarized
waves of either opposite handedness or opposite directions of
incidence. Fig.~\ref{fig-TRA}a presents the absolute difference in
total transmission of RCP and LCP waves, $\Delta T=T_{+}-T_{-}$,
plotted for different directions of incidence. The difference is
resonant in a narrow range of frequencies from 5.5 to 5.9~GHz
reaching a maximum of about 15~\% near 5.7~GHz. The plot shows that
for circularly polarized waves incident on the front of the
structure the total transmission for RCP is substantially higher
than for LCP, while for waves incident on the back the situation is
reversed and the metamaterial appears to be more transparent to LCP.
A similar polarization and directional asymmetry can be seen in
reflection: the total reflectivity (i.e. reflectivity that would be
measured with a polarization insensitive detector) for RCP-waves
incident on the metamaterial's front is larger than that for
LCP-waves, and the reflectivity difference changes sign upon
reversal of the propagation direction (see Fig.~\ref{fig-TRA}b).

Fig.~\ref{fig-TRA}c shows the asymmetric behavior of the
metamaterial in terms of losses. Here we present the absolute
difference in metamaterial absorption of RCP and LCP calculated as
$\Delta A=A_{+}-A_{-}$, where $A_{\pm}=1-T_{\pm}-R_{\pm}$. The
structure appeared to be dichroic with respect to the incident
circular polarization, but more importantly its dichroic response
also depended on the direction of propagation, exhibiting asymmetry
very similar to that of total transmission and reflection. Since at
microwave frequencies copper is a very good conductor, while the
metamaterial grating did not diffract below 20~GHz, the losses must
have resulted from absorption in the dielectric substrate
($\textrm{Im}~\varepsilon \simeq 0.2$).

\begin{figure}[t!]
\includegraphics[width=85mm]{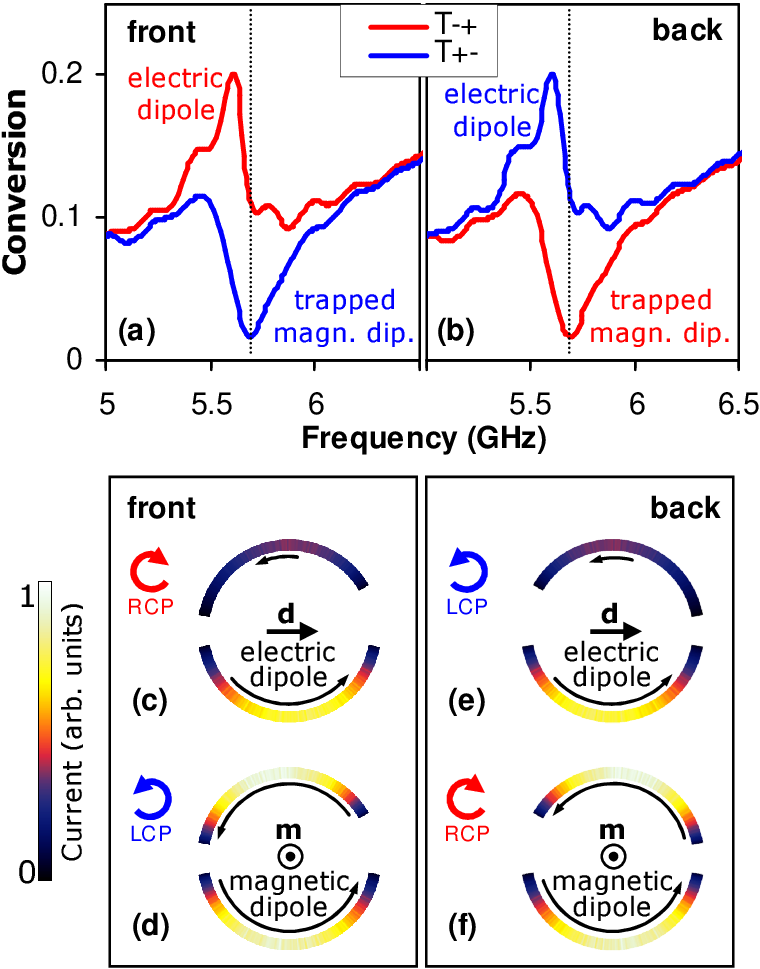}
\caption{\label{fig-conversion}(Color online) Resonant circular
conversion dichroism. Panels (a) and (b) show polarization
conversion spectra measured in transmission for RCP (red) and LCP
(blue) microwaves normally incident on the structure's front and
back. Panels (c)-(f) present the magnitude of resonant currents
along the ring at about 5.7~GHz. The instantaneous direction of the
currents is indicated by arrows: (c) RCP incident on the structure's
front excites a strongly scattering electric dipole-like current
mode (strong polarization conversion). (d) LCP excites a
weakly-scattering magnetic mode (conversion minimum). Panels (e) and
(f) show current distributions for circularly polarized waves
incident on the structure from the opposite side.}
\end{figure}

Within the accuracy of our measurements, we found that the direct
transmission of circular polarization was identical for opposite
polarization states, i.e. $t_{++} = t_{--}$, as well as for opposite
directions of propagation. The same applied to the reflected
unconverted polarization components $r_{-+}$ and $r_{+-}$. Thus the
asymmetric response must have been completely controlled by
polarization conversion. This is illustrated by Fig.
\ref{fig-conversion}a, which shows the polarization conversion
levels for LCP and RCP incident on the metamaterial's front. Within
experimental accuracy the efficiencies of conversion in transmission
and reflection are the same, i.e. $T_{-+}=R_{++}$ and
$T_{+-}=R_{--}$. At around 5.7~GHz the polarization conversion for
RCP reaches 20~\%, while the conversion for LCP drops below 2~\%,
leading to a very large circular conversion dichroism within a
narrow spectral range. The situation is similar for waves incident
on the back of the metamaterial, however, with reversed roles of RCP
and LCP.

To understand the nature of the large circular conversion dichroism
we numerically calculated the distribution of currents in the split
rings excited by circularly polarized waves normally incident on the
metamaterial using a full 3D Maxwell FEM solver in the frequency
domain (see Fig.~\ref{fig-conversion}c-f). The pattern of the
metamaterial was modeled as an array of ideally conducting metal
split rings of zero thickness (which is a fair approximation at
microwave frequencies), while all other parameters of the structure
were chosen identical to those of the real sample. Our modeling took
advantage of the periodicity of the structure, which was represented
by a single unit cell with periodic boundary conditions imposed on
the computational domain in the lateral directions. We found that
there are two distinct regimes of \emph{resonant} excitation which
depend on the handedness of the incident wave and the direction in
which the wave enters the structure. The response to excitation with
a right circularly polarized wave (RCP, $+$) incident on the front
of the structure is essentially electric dipolar in nature with the
dipole oriented along the split of the ring (Fig.
\ref{fig-conversion}c). Radiation of the induced oscillating linear
dipole can be presented as a sum of left and right circular
polarizations where the left-handed component of scattering gives
rise to strong resonant polarization conversion (red curve, Fig.
\ref{fig-conversion}a). On the contrary, the response to excitation
with a left circularly polarized wave (LCP, $-$) is essentially
magnetic dipolar. Here the induced magnetic moment is perpendicular
to the metamaterial plane and is created by anti-symmetric currents
flowing in opposite sectors of the ring (see Fig.
\ref{fig-conversion}d). The anti-symmetric current mode is weakly
coupled to free space, scattering is low
\cite{PRL_Fedotov_2007_TrappedModes} and polarization conversion is
at its minimum (blue curve, Fig. \ref{fig-conversion}a). Due to weak
scattering, energy coupled to the magnetic mode is trapped in the
anti-symmetric current oscillation and eventually dissipated in the
lossy dielectric substrate, which results in large absorption losses
(Fig. \ref{fig-TRA}c). When the propagation direction is reversed,
i.e. the wave enters the structure from the opposite side, the
perceived sense of planar chirality of the design reverses: the
larger split on the right now appears to be on the left of the ring.
Now the roles of left and right circular polarizations are swapped
around: the left circular polarization excites an electric response
(Fig. \ref{fig-conversion}e) while the right circular polarization
excites a predominantly magnetic response in the metamaterial (Fig.
\ref{fig-conversion}f). Indeed, polarization conversion is now at
its maximum for left circular polarization and at a minimum for
right circular polarization, as shown in Fig. \ref{fig-conversion}b.

Thus the asymmetric phenomena arise from excitation of an electric
dipole-like ``conversion" mode and a weakly scattering
anti-symmetric ``absorption" mode by opposite circular
polarizations. Note that this microscopic mechanism should in
general apply to all planar metamaterials exhibiting asymmetric
transmission/reflection.

\begin{figure}[t!]
\includegraphics[width=85mm]{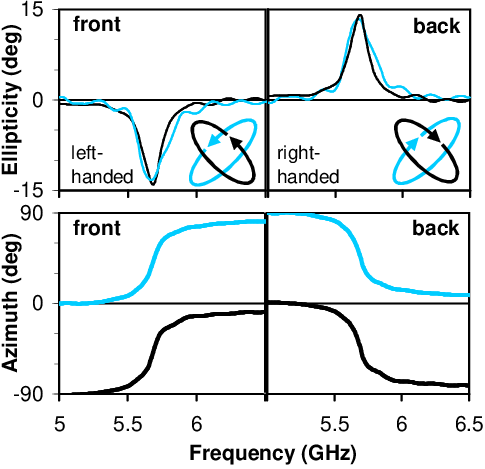}
\caption{\label{fig-eigenstates}(Color online) Dispersions of
ellipticity angle and azimuth of both of the structure's
transmission eigenstates represented by blue and black lines
correspondingly.}
\end{figure}

Although the structure shows strong resonant polarization conversion
for circularly polarized electromagnetic waves, certain polarization
states remain unchanged on transmission. Fig. \ref{fig-eigenstates}
shows the ellipticity angle and azimuth of the metamaterial's
transmission eigenstates. In non-resonant regions the eigenstates
are a pair of orthogonal linear polarizations with the azimuth
corresponding approximately to the directions along and
perpendicular to the ring's split. At the resonance, however, the
eigenstates become co-rotating orthogonal ellipses. The reflection
eigenstates are the same as the transmission eigenstates, while for
the eigenstates of the opposite propagation direction the handedness
is reversed.

In conclusion, we have demonstrated strong resonant directional and
polarization asymmetry of normal incidence reflection and
transmission of circularly polarized light from and through a planar
chiral metamaterial. Given that nano-scaled versions of the
metamaterial structure are naturally suited for the existing planar
fabrication technologies it may have wide potential applications in
photonic devices that exploit direction/polarization-dependent
operation, such as asymmetric wave splitters and circulators.

\begin{acknowledgments}
Financial support of the Engineering and Physical Sciences Research
Council (UK) through the Nanophotonics Portfolio Grant and a CA
Fellowship (VAF) is acknowledged.
\end{acknowledgments}



\begin{thebibliography}{10}
\expandafter\ifx\csname
natexlab\endcsname\relax\def\natexlab#1{#1}\fi
\expandafter\ifx\csname bibnamefont\endcsname\relax
  \def\bibnamefont#1{#1}\fi
\expandafter\ifx\csname bibfnamefont\endcsname\relax
  \def\bibfnamefont#1{#1}\fi
\expandafter\ifx\csname citenamefont\endcsname\relax
  \def\citenamefont#1{#1}\fi
\expandafter\ifx\csname url\endcsname\relax
  \def\url#1{\texttt{#1}}\fi
\expandafter\ifx\csname urlprefix\endcsname\relax\def\urlprefix{URL
}\fi \providecommand{\bibinfo}[2]{#2}
\providecommand{\eprint}[2][]{\url{#2}}

\bibitem[{\citenamefont{Fedotov et~al.}(2006)\citenamefont{Fedotov, Mladyonov,
  Prosvirnin, Rogacheva, Chen, and
  Zheludev}}]{PRL_Fedotov_2006_AsymmetricTransmissionMW}
\bibinfo{author}{\bibfnamefont{V.~A.} \bibnamefont{Fedotov}},
  \bibinfo{author}{\bibfnamefont{P.~L.} \bibnamefont{Mladyonov}},
  \bibinfo{author}{\bibfnamefont{S.~L.} \bibnamefont{Prosvirnin}},
  \bibinfo{author}{\bibfnamefont{A.}~\bibnamefont{Rogacheva}},
  \bibinfo{author}{\bibfnamefont{Y.}~\bibnamefont{Chen}}, \bibnamefont{and}
  \bibinfo{author}{\bibfnamefont{N.~I.} \bibnamefont{Zheludev}},
  \bibinfo{journal}{Phys. Rev. Lett.} \textbf{\bibinfo{volume}{97}},
  \bibinfo{pages}{167401} (\bibinfo{year}{2006}).

\bibitem[{\citenamefont{Fedotov
  et~al.}(2007{\natexlab{a}})\citenamefont{Fedotov, Schwanecke, Zheludev,
  Khardikov, and Prosvirnin}}]{NanoLett_Fedotov_2007_AsymmetricTransmission}
\bibinfo{author}{\bibfnamefont{V.~A.} \bibnamefont{Fedotov}},
  \bibinfo{author}{\bibfnamefont{A.~S.} \bibnamefont{Schwanecke}},
  \bibinfo{author}{\bibfnamefont{N.~I.} \bibnamefont{Zheludev}},
  \bibinfo{author}{\bibfnamefont{V.~V.} \bibnamefont{Khardikov}},
  \bibnamefont{and} \bibinfo{author}{\bibfnamefont{S.~L.}
  \bibnamefont{Prosvirnin}}, \bibinfo{journal}{Nano Lett.}
  \textbf{\bibinfo{volume}{7}}, \bibinfo{pages}{1996}
  (\bibinfo{year}{2007}{\natexlab{a}}).

\bibitem[{\citenamefont{Schwanecke et~al.}(2008)\citenamefont{Schwanecke,
  Fedotov, Khardikov, Prosvirnin, Chen, and
  Zheludev}}]{NanoLett_Schwanecke_2008_AsymTrans}
\bibinfo{author}{\bibfnamefont{A.~S.} \bibnamefont{Schwanecke}},
  \bibinfo{author}{\bibfnamefont{V.~A.} \bibnamefont{Fedotov}},
  \bibinfo{author}{\bibfnamefont{V.~V.} \bibnamefont{Khardikov}},
  \bibinfo{author}{\bibfnamefont{S.~L.} \bibnamefont{Prosvirnin}},
  \bibinfo{author}{\bibfnamefont{Y.}~\bibnamefont{Chen}}, \bibnamefont{and}
  \bibinfo{author}{\bibfnamefont{N.~I.} \bibnamefont{Zheludev}},
  \bibinfo{journal}{Nano Lett.} \textbf{\bibinfo{volume}{8}},
  \bibinfo{pages}{2940} (\bibinfo{year}{2008}).

\bibitem[{\citenamefont{Drezet et~al.}(2008)\citenamefont{Drezet, Genet,
  Laluet, and Ebbesen}}]{OptExp_Ebbesen_2008_AsymTrans}
\bibinfo{author}{\bibfnamefont{A.}~\bibnamefont{Drezet}},
  \bibinfo{author}{\bibfnamefont{C.}~\bibnamefont{Genet}},
  \bibinfo{author}{\bibfnamefont{J.-Y.} \bibnamefont{Laluet}},
  \bibnamefont{and} \bibinfo{author}{\bibfnamefont{T.~W.}
  \bibnamefont{Ebbesen}}, \bibinfo{journal}{Opt. Exp.}
  \textbf{\bibinfo{volume}{16}}, \bibinfo{pages}{12559} (\bibinfo{year}{2008}).

\bibitem[{\citenamefont{Zhukovsky et~al.}(2008)\citenamefont{Zhukovsky,
  Galynsky, and Novitsky}}]{MetaCongress_Zhukovsky_2008_EllDichroism}
\bibinfo{author}{\bibfnamefont{S.~V.} \bibnamefont{Zhukovsky}},
  \bibinfo{author}{\bibfnamefont{V.~M.} \bibnamefont{Galynsky}},
  \bibnamefont{and} \bibinfo{author}{\bibfnamefont{A.~V.}
  \bibnamefont{Novitsky}}, in \emph{\bibinfo{booktitle}{Metamaterials Congress,
  Pamplona, Spain}} (\bibinfo{year}{2008}).

\bibitem[{\citenamefont{Kopp et~al.}(2004)\citenamefont{Kopp, Churikov, Singer,
  Chao, Neugroschl, and Genack}}]{Science_Genack_ChiralFiberGratings}
\bibinfo{author}{\bibfnamefont{V.~I.} \bibnamefont{Kopp}},
  \bibinfo{author}{\bibfnamefont{V.~M.} \bibnamefont{Churikov}},
  \bibinfo{author}{\bibfnamefont{J.}~\bibnamefont{Singer}},
  \bibinfo{author}{\bibfnamefont{N.}~\bibnamefont{Chao}},
  \bibinfo{author}{\bibfnamefont{D.}~\bibnamefont{Neugroschl}},
  \bibnamefont{and} \bibinfo{author}{\bibfnamefont{A.~Z.}
  \bibnamefont{Genack}}, \bibinfo{journal}{Science}
  \textbf{\bibinfo{volume}{305}}, \bibinfo{pages}{74} (\bibinfo{year}{2004}).

\bibitem[{\citenamefont{Shvets.}(2006)}]{APL_Shvets_2006_ChiralWaveguide}
\bibinfo{author}{\bibfnamefont{G.}~\bibnamefont{Shvets.}},
  \bibinfo{journal}{Appl. Phys. Lett.} \textbf{\bibinfo{volume}{89}},
  \bibinfo{pages}{141127} (\bibinfo{year}{2006}).

\bibitem[{\citenamefont{Hwang et~al.}(2005)\citenamefont{Hwang, Song, Park,
  Nishimura, Toyooka, Wu, Takanishi, Ishikawa, and Takezoe}}]{hwang-2005-eta}
\bibinfo{author}{\bibfnamefont{J.}~\bibnamefont{Hwang}},
  \bibinfo{author}{\bibfnamefont{M.~H.} \bibnamefont{Song}},
  \bibinfo{author}{\bibfnamefont{B.~Y.} \bibnamefont{Park}},
  \bibinfo{author}{\bibfnamefont{S.}~\bibnamefont{Nishimura}},
  \bibinfo{author}{\bibfnamefont{T.}~\bibnamefont{Toyooka}},
  \bibinfo{author}{\bibfnamefont{J.~W.} \bibnamefont{Wu}},
  \bibinfo{author}{\bibfnamefont{Y.}~\bibnamefont{Takanishi}},
  \bibinfo{author}{\bibfnamefont{K.}~\bibnamefont{Ishikawa}}, \bibnamefont{and}
  \bibinfo{author}{\bibfnamefont{H.}~\bibnamefont{Takezoe}},
  \bibinfo{journal}{Nature Mat.} \textbf{\bibinfo{volume}{4}},
  \bibinfo{pages}{383} (\bibinfo{year}{2005}).

\bibitem[{\citenamefont{Fedotov
  et~al.}(2007{\natexlab{b}})\citenamefont{Fedotov, Rose, Prosvirnin,
  Papasimakis, and Zheludev}}]{PRL_Fedotov_2007_TrappedModes}
\bibinfo{author}{\bibfnamefont{V.~A.} \bibnamefont{Fedotov}},
  \bibinfo{author}{\bibfnamefont{M.}~\bibnamefont{Rose}},
  \bibinfo{author}{\bibfnamefont{S.~L.} \bibnamefont{Prosvirnin}},
  \bibinfo{author}{\bibfnamefont{N.}~\bibnamefont{Papasimakis}},
  \bibnamefont{and} \bibinfo{author}{\bibfnamefont{N.~I.}
  \bibnamefont{Zheludev}}, \bibinfo{journal}{Phys. Rev. Lett.}
  \textbf{\bibinfo{volume}{99}}, \bibinfo{pages}{147401}
  (\bibinfo{year}{2007}{\natexlab{b}}).

\bibitem[{\citenamefont{Papasimakis et~al.}(2008)\citenamefont{Papasimakis,
  Fedotov, Fu, Tsai, and Zheludev}}]{arXiv_Papasimakis_2008_Coherent}
\bibinfo{author}{\bibfnamefont{N.}~\bibnamefont{Papasimakis}},
  \bibinfo{author}{\bibfnamefont{V.~A.} \bibnamefont{Fedotov}},
  \bibinfo{author}{\bibfnamefont{Y.~H.} \bibnamefont{Fu}},
  \bibinfo{author}{\bibfnamefont{D.~P.} \bibnamefont{Tsai}}, \bibnamefont{and}
  \bibinfo{author}{\bibfnamefont{N.~I.} \bibnamefont{Zheludev}},
  \bibinfo{journal}{arXiv.org, 0809.2361v1}  (\bibinfo{year}{2008}).

\end{thebibliography}
\end{document}